\documentclass[12 pt]{amsart}
\usepackage{graphicx}
\usepackage{verbatim}
\usepackage{comment}
\usepackage{calc}
\usepackage{color}
\usepackage{soul}
\setul{}{2pt}
\usepackage{tkz-euclide}
\usetkzobj{all}
\usepackage{placeins} 
\usepackage[top=1.5in, bottom=1.5in, left=1in, right=1in]{geometry}
\usepackage{setspace}
\begin{document}

\title{Discrete power law with exponential cutoff and Lotka's Law}

\author{Lawrence Smolinsky}
\address{Department of Mathematics\\
Louisiana State University\\
Baton Rouge, LA 70803, USA}
\email{smolinsk@math.lsu.edu}

\date{\today}

\begin{abstract}  The first bibliometric law appeared in Alfred J. Lotka's 1926 examination of author productivity in chemistry and physics.  The result is that the productivity distribution is thought to be described by a power law.  In this paper, Lotka's original data on author productivity in chemistry is reconsidered by comparing the fit of the data to both a discrete power law and a discrete power law with exponential cutoff. 
\end{abstract}

\maketitle
\pagestyle{myheadings}
\markboth{L. Smolinsky}{Discrete power law with exponential cutoff}

\section{Introduction.}  According to Dorothy H. Hertzel (2011 Table 2 p. 559), the first seminal papers in bibliometrics started with statistical studies by Cole and Eales in 1917 and Hulme and Wyndham in 1922.  William Hood and Concepci\'{o}n Wilson (2001 p. 295) point out that the next step, ``the
discovery of certain regularities, distributions or laws," was of fundamental importance to the development of the fields of bibliometrics, scientometrics, and informetrics.  The earliest of these laws was Lotka's.  In 1926 Alfred J. Lotka published his study on the publication rates of authors in chemistry (Lotka 1926).\footnote{Lotka also included a smaller study of physics authors, which we do not address.}  He proposed an inverse square relationship that has since been generalized to a power law.\footnote{In fact, Lotka used the generalize Lotka's law of Eq.~(1) with $\alpha=1.888$ and $1/\zeta(1.888)=0.5669$ for his computations in column 9 of Table 1 (Lotka 1926) and with $\alpha=2$ and $1/\zeta(2)=0.6079$ in column 11 of Table 1.}  This form of Lotka's law is that that for some $\alpha$, the percentage of senior authors who are the senior author of $n$ works is
\begin{equation}
\label{eq:lotkapowerlaw}
p_0(x) =\frac {x^{-\alpha}} {\zeta{(\alpha)}}\qquad \text{ for each } x\in \mathbb N,
\end{equation}
where $\zeta(\alpha)$ is the Riemann zeta function.  One might allow different $\alpha$'s for different fields.

Lotka gathered data from Chemical Abstracts 1907-1916 including 6891 senior authors.  He examined the curve of the number of works (abscissa) versus the percentage of authors having that number of works (ordinate) to give 66 data points ($x_i,y_i)$ for $i= 1,\cdots, 66$ to approximate the (discrete) curve.  A second philosophy is to consider the 6891 pieces of data ($z_i$ for $i= 1,\cdots,6891$) as independent sample points from the probability distribution of a random variable $Z$, where
$$Z(\text{senior author}) = \text{\# of papers on which he or she is the senior author}.$$
The first viewpoint lends itself to curve fitting techniques while the second to maximum likelihood estimates of the probability mass function of $Z$.

Lotka's original approach was to truncate his data and take the logarithm of both the number of works variable $x$ and the percentage of authors variable $y$ and perform a linear least squares fit.  A good summary is in Coile (1977).  Since at least Paul Travis Nicholls's works (1987, 1989), the method of the maximum likelihood estimator with the Kolmogorov-Smirnov test has been viewed as superior.  Ronald Rousseau (2002 p. 319), in his discussion of informetric research on power laws,  adds, ``most large datasets usually cannot be described by a Lotka distribution (in a statistical sense, using these [Kolmogorov-Smirnov or chi-squared] goodness-of-fit tests)."  These viewpoints have more recently been championed in the influential article by Clauset, Shalizi, and Newman (2009).  Clauset et al. add to the discussion a comparison of power laws with other long tailed distributions using a log likelihood ratio test, and they included the continuous power law with exponential cutoff.

In this paper, Lotka's original data is examined, and the discrete power law is compared to a discrete power law with exponential cutoff via a log likelihood ratio test.  The discrete power law is a special case of the discrete power law with exponential cutoff.  Lotka's law has been well studied.   Nicholls (1987 p. 23-24) gives a list of studies that includes 20 proposed alternative distributions to the discrete power law.  We begin with a review of various methods and results to estimate $\alpha$ in Eq.~(\ref{eq:lotkapowerlaw}).  Linear least square estimates were computed using excel and all other computations were done using Mathematica 10.0.0.0.

\section{Acknowledgements.}  I am grateful to Stephen Bensman for suggesting I look at Lotka's original paper and to Luis Escobar for guidance to the statistics references.

\section{Discrete power law estimates}  We examine power law estimates on Lotka's full data set (untruncated data) for comparison with the power law with exponential cutoff.  Lotka's approach was to organize his data into the number of works variable $x$ and the percentage of authors variable $y$, truncate his data points, take the logarithm of both variables, and perform a linear least squares fit.  This method allows for two variables in the line of best fit:  The slope and the intercept in the equation
\begin{equation}
\label{eq:log-logLLS}
\ln(y(x)) = -\alpha \ln(x) + b
\end{equation}
which corresponds to the power law expression
\begin{equation}
\label{eq:twovarpowerlaw}
y(x)=e^b/x^\alpha.
\end{equation} 
Performing linear least squares fit on Lotka's truncated data set, the results are $\alpha = -1.8877$ and $b= -0.5349$ or $e^b=0.5857$.  Lotka only used least squares to determine $\alpha$ and then choose the constant $e^b$ so as to arrive at a probability distribution.  Lotka's $\alpha$ is $1.888$, and his constant is $ 1/\zeta(1.888) =0.5669$.  There are issues both with truncating the data (Nicholls 1987, 1989; Bensman 2015) and with taking logarithms followed by linear least squares (Nicholls 1987, Clauset et al. 2009).

One issue with least squares fit to a power function is that $\alpha$ and $b$ found for Eq.~(\ref{eq:log-logLLS}) or Eq.~(\ref{eq:twovarpowerlaw}), does not necessarily define a probability distribution.  We should remark that those who use the method of least squares are not naive to think that is does.  Rather, many (including Lotka) employ separate methods to estimate $b$ and $\alpha$ (Nicholls 1987).  To fit the best line that corresponds to a discrete probability distribution using the log-log linear least squares method, one would restrict to lines of the form 
\begin{equation}
\label{eq:linearleastsquaresprob}
\ln(y(x)) = -\alpha \ln(x) - \ln\zeta(\alpha ),
\end{equation}
which is now a single variable minimization problem.  This method is constrained linear least squares in Table~(\ref{table:powerlaw}). 

A second issue is that the standard notion of ``best fit" by Gauss and Legendre to fit the data (i.e, measuring best by the sum of the squares of residuals) does not apply when logarithms are applied first.  We apply non-linear least squares directly on Eq.~(\ref{eq:twovarpowerlaw}) and on the family Eq.~(\ref{eq:lotkapowerlaw}), i.e., constrained non-linear least squares, in Table~(\ref{table:powerlaw}).  However, either way, an $r^2$ calculation is not informative since normality of residuals cannot be assumed and $r^2$ is weakly correlated to goodness-of-fit for Lotka's distribution (Nicholls, 1987, 1989).

Table~(\ref{table:powerlaw}) gives the results of calculations performed in Mathematica using: linear least squares on Eq.~(\ref{eq:log-logLLS});  constrained linear least squares restricting to lines of the form of Eq.~(\ref{eq:linearleastsquaresprob}); non-linear least squares on the two variable family Eq.~(\ref{eq:twovarpowerlaw}); and constrained non-linear least squares restricting to the one variable family Eq.~(\ref{eq:lotkapowerlaw}).

We next examine the determination of $\alpha$ in Eq.~(\ref{eq:lotkapowerlaw}) by use of the maximum likelihood estimator.  This method has its flaws too.  Luong and Doray (1996) point out it is efficient, but it is not robust, sensitive to outliers. Its score function depends on the sample mean, so its influence function is unbounded (Huber 1981).  This method assumes that the 6891 pieces of data ($z_i$ for $i= 1,\cdots,6891$) are independent samples of the random variable $Z$ in the introduction, which has probability mass function Eq.~(\ref{eq:lotkapowerlaw}).  Performing a maximum likelihood estimate on Lotka's full data set and the family of discrete power laws, the result is $\alpha = 1.9665$ and the log likelihood is $-11705.1$ (Tables \ref{table:powerlaw} and \ref{table:exponentialcutoff}).

\begin{center}
\begin{table}
\caption{Power law estimates on Lotka's full data set.}
\begin{tabular} {  |  c   |  c  |  c  | }
 \hline 
Method & $\alpha$   & $b$ \\ \hline\hline
linear least squares & 1.8122 & - 0.8812  \\ \hline
constrained linear least squares & 1.8985 & ---  \\ \hline
non-linear least squares  & 1.9018 & -0.5466  \\ \hline 
constrained non-linear least squares  & 1.9185 & ---  \\ \hline
maximum likelihood  & 1.9665 &  ---  \\ \hline
\end{tabular}
\label{table:powerlaw}
\end{table}
\end{center}

\section{Discrete power law with exponential cutoff.}

A more general model that includes the discrete power law family is a discrete power law with exponential cutoff.  It has two parameters $\alpha$ and $\beta$.  The definition of the probability mass function involves the polylogarithm also known as Jonqui\'{e}re's function.\footnote{Sometimes this function is called the $\alpha$-polylogarithm or fractional polylogarithm and the term polylogarithm is reserved for the case $\alpha$ an integer.}  A short history of the function is given in the introduction to Costin and Garoufalidis (2009) and a short list of its properties is in Lee (1997).  The polylogarithm is a function defined for all complex $s$ and $z$, and it satisfies
$$\text {Li}_{s}(z) =  \sum_{n=1}^\infty \frac {z^n}{n^s} \text{ if } |z|<1 \quad \text {and } \quad 
\text {Li}_{s}(1) =  \zeta(s) \text{ if } \text{Re } s>1$$
Define the probability mass function of a discrete power law with exponential cutoff as:
\begin{equation}
\label{eq:exponentialcutoff}
p_a(x) = \frac 1{\text {Li}_{\alpha}(e^{\beta})} \frac {e^{\beta x}}{x^\alpha} \ \ \ \ \ x\in \mathbb N.
\end{equation}
The space of parameters for this probability mass function is
$$\{ (\alpha,\beta)\in \mathbb R^2 \mid   \beta < 0 \text{ or }  \beta = 0 \text{ and }  \alpha > 1 \}$$
If $\beta = 0$ and  $\alpha > 1$, then the discrete power law with exponential cutoff specializes to the discrete power law.

The likelihood function for Lotka's data $\mathbf z = (z_1,\cdots, z_{6891})$ is 
$$\mathcal{L}(\alpha,\beta | \mathbf z) =\prod_{i=1}^{6891}\frac 1{\text {Li}_{\alpha}(e^{\beta})} \frac {e^{\beta z_i}}{z_i^\alpha}$$
The functions $\mathcal{L}$ and  $\text {Li}_{\alpha}(e^{\beta})$ are infinitely differentiable when $\beta < 0$ but have problems on part of the boundary (i.e., $\beta = 0$) where the probability mass function is the discrete power law family.  When $\beta=0$ the probability mass function may not have finite variance either.  

Finding the maximum likelihood for Lotka's full data set and the discrete power law with exponential cutoff distribution, Eq.~(\ref{eq:exponentialcutoff}), the result is $\alpha_a= 1.81524$ and $\beta_a = -0.0172868$ with a log likelihood of $-11646.2$.

To compare the log likelihoods, we use the log likelihood ratio test.  The test statistic, which is denoted $-2\ln (\lambda)$, is:
 $$-2 (\text {log likelihood of the null hypothesis} - \text {log likelihood of the alternative hypothesis}).$$  
 We apply the log likelihood ratio test on nested models (Boos and Stefanski 2013).  One model, the alternative hypothesis, is the two-parameter family of the discrete power law with exponential cutoff.  Second model, null hypothesis, will be the restriction to $\beta=0$ or to a particular small $\beta$, both possible null hypotheses have one parameter.  The test we will apply has a simple hypothesis (Wilks 1936) meaning the null hypothesis is just a restriction of the parameter space.  Under further suitable hypotheses, the test statistic has an approximate chi-squared distribution with one-degree of freedom.  It is one-degree of freedom since one model has one parameter and the second model has two parameters.  The alternative hypothesis encompasses the null hypothesis and hence must fit better than null hypothesis.  However, the question is does it fit better with statistical significance.
 
We caution the reader that Wilks's theorem does not justify a chi-squared statistic interpretation using null hypothesis $p_0$.   Hence, we consider a comparison with second null hypothesis, $p_1$, obtained by moving $\beta$ slightly into the interior of region in which Wilks's theorem applies.  Consider the model with $\beta = - 0.000001$.   This one parameter model has probability mass function
$$p_1(x) = \frac 1{\text {Li}_{\alpha}(e^{- 0.000001})} \frac {e^{- 0.000001 x}}{x^\alpha} \ \ \ \ \ x\in \mathbb N.$$
Performing a maximum likelihood estimate to obtain parameters for the $p_1$ model yields $\alpha_1= 1.96643$ and $\beta = - 0.000001$.  This probability mass function is virtually indistinguishable from $p_0$ 
on the range where there is data (from 1 to 346).  The difference between the curves $|p_0-p_1|$ is less than 0.000023 and the ratio $p_0/p_1$ varies between 0.9997598 and 1.0000389.  Models are imperfect by their statistical nature and the region on which data exists determines them.  These two models, $p_0$ and $p_1$, are identical within reasonable rounding.  We may compare $p_a$ and $p_1$ in order to avoid issues with the hypotheses of Wilks's theorem and allow a chi-squared statistic interpretation.

Results are given in Table~(\ref{table:exponentialcutoff}).  The test yields $-2\ln (\lambda) = 117.8$.  The critical significance level for $\chi^2$ with one degree of freedom is $P(\chi^2\le 0.99) = 6.64$.  We can conclude that the fit is significantly worse under the null hypothesis.  We also include a two-sided Kolmogorov-Smirnov goodness of fit test.  The test statistic is the supremum (here just the maximum) of the distance between the cumulative distribution function of the data from the theoretical cumulative distribution function, i.e., the distance in the $l$-infinity norm.  The significance level for 95\% in the continuous approximation is $\frac{1.36}{\sqrt{6891}} \approx 0.01638$.  This value is conservative according to Conover (1999).\footnote{Conover gives recursive formulas to compute the exact p-value for discrete distributions, but it is beyond the author's computing power.}  The test seems to indicate acceptance for $p_a$ and rejection of $p_0$, however, $p_0$ may actually be within acceptance if the actual p-value could be computed.

\begin{center}
\begin{table}
\caption{Log likelihood comparisons}
\begin{tabular} {  |  c   |  c  |  c  | c  |  c  |}
 \hline 
hypothesis &   \parbox[t]{4 cm} {parameters\\using max likelihood}  &log likelihood  & $-2\ln (\lambda)$ & KS D-statistic \\ \hline\hline
 \parbox[t]{5.5 cm}{alternate\\general exponential cutoff ($p_a$)}  &  \parbox[t]{4 cm}{$\alpha_a= 1.8152375$\\$\beta_a =-0.0172869$}  & -11646.153 & & 0.007589 \\ \hline
 \parbox[t]{5.5 cm}{$\beta =0$\\discrete power law ($p_0$)}  &  \parbox[t]{4 cm}{$\alpha_0= 1.9665088$\\$\beta =0$}  &   -11705.124 & 117.832 & 0.016968  \\ \hline
  \parbox[t]{5.5 cm}{$\beta =-0.000001$\\virtual discrete power law  ($p_1$)} & \parbox[t]{4 cm}{$\alpha_1= 1.9664271$\\$\beta =- 0.0000010$}& -11705.068 & 117.941  & 0.016945\\ \hline
\end{tabular}
\label{table:exponentialcutoff}
\end{table}
\end{center}

\FloatBarrier

\section{Conclusion.}

The log likelihood ratio statistic $-2\ln (\lambda) = 117.8$ is large and the statistical preference is for the power law with exponential cutoff (as is the preference from the Kolmogorov-Smirnov test).   One should be cautious and note that with large data samples (here more than 6000 data points) small differences in probability distributions can make a large difference in a statistic.  Such statistics are more convincing when samples are truly drawn as repeated trials from an ideal distribution, but Lotka's law is more likely a result of some compound Poisson process (e.g., Huber 2002) whose details we would still like to understand.  Replacing assumptions in a Poisson process such as career longevity and production rates with on average assumptions or with approximate distributions drawn from statistics on authors will change the ultimate resultant distribution.

The Kolmogorov-Smirnov statistic measures the distance between the data cumulative distribution function and the theoretical cumulative distribution functions.  The statistic is much less for $p_a$ than $p_0$.  One might think the difference is a reflection of the difference in the tail of the two distributions, but it is worth noting that the maximum vertical distance between the theoretical distributions and the data distribution actually occurs at $x=1$ for the three probability mass functions.  This is typical since the vertical height of tail is small whether it is a ``fat tail" or a ``thin tail."

We close with an observation about Lotka's method of using logarithms, truncation, and linear least squares.  In an initial reading of Lotka's paper, it appears that Lotka was very careful not to destroy the information that there is a long tail even though he truncated his data.  This appearance comes from the fact that Lotka 

(1) computed the data probability distribution and then truncated the tail, rather than 

(2) truncate the data first and then compute a finite range probability distribution.\\
Lotka's probabilities do not sum to one in his table, because he truncated the distribution (1) rather than the data (2).

If one was to test the data distribution using a something like a Kolmogorov-Smirnov test, this care might make a difference since (1) and (2) are different distributions.  However, when one takes logarithms of (1) and (2), one obtains parallel curves.  The lines of best fit after taking logarithms of (1) and (2) will have the same slope and only differ in the intercepts!  It is simply the rules of logarithms as (1) and (2) differ from the raw data only by having different multiplicative normalization constants.  Since Lotka only uses the slope of the line of best fit, his caution does not make a difference.

$$\text {References}$$

\noindent
Bensman, SJ, (2015) forthcoming.

\vspace {2 mm}

\noindent
Boos, D.D. and Stefanski, L.A. (2013).  \textit {Essential Statistical Inference:  Theory and Methods.} Springer texts in Statistics.  New York: Springer.

\vspace {2 mm}

\noindent
Clauset, A., Shalizi, C.R., and Newman, M.E.J. (2009).  Power-Law Distributions in Empirical Data.  \textit{SIAM Review}, 51(4), 661-703.

\vspace {2 mm}

\noindent
Conover, W.J. (1999). \textit {Practical nonparametric statistics.} (3d ed.). New York: Wiley. 

\vspace {2 mm}

\noindent
Costin, O. and Stavros Garoufalidis, S. (2009).  Resurgence of the fractional polylogarithms.  \textit{Mathematical Research Letters} 16 (5), 817-826.

\vspace {2 mm}

\noindent
Coile, R.C. (1977) Lotka's Frequency Distribution of Scientific Productivity.
\textit{Journal of the American Society for Information Science}, 28(6) 366-370.

\vspace {2 mm}

\noindent
Hertzel, D.H. (2011). Bibliometric Research: History [ELIS Classic] In
\textit{Encyclopedia of Library and Information Sciences}, Third Edition. Taylor and Francis: New York,
Published online: 11 Oct 2011; 546-583.

\vspace {2 mm}

\noindent
Hood, W.W., Wilson, C.S. (2001).  The literature of bibliometrics, scientometrics, and informetrics. \textit{Scientometrics}, 52 (2), 291-314.

\vspace {2 mm}

\noindent
Lee, M.H. (1997).  Polylogarithms and Riemann's zeta function. \textit{Physical Review E}, 56 (4), 3909?3912

\vspace {2 mm}

\noindent
Lotka, A.J. (1926).  The frequency distribution of of scientific productivity.
\textit{Journal of the Washington Academy of Sciences}, 16 (12), 317-323.

\noindent
Luong, A., Doray, L.G. Goodness of fit test statistics for the zeta family.
\textit{Insurance: Mathematics and Economics}, 19, 45-53.

\vspace {2 mm}

\noindent
Nicholls, P. T. (1987). \textit{The Lotka hypothesis and bibliometric methodology},
Ph.D. Thesis. School of Library and Information Science, The University of Western Ontario, London, Canada.
Digitized Theses. Paper 1578. http://ir.lib.uwo.ca/digitizedtheses/1578

\vspace {2 mm}

\noindent
Nicholls, P.T. (1989).  Bibliometric Modeling Processes and the Empirical Validity of Lotka's Law.
\textit{Journal of the American Society for Information Science and Technology},  40(6), 379-385.

\vspace {2 mm}

\noindent
Rousseau, R. (2002) Lack of standardisation in informetric research.  Comments on ``Power laws of research output.  Evidence for journals of economics" by Matthias Sutter and Martin G. Kocher. \textit{Scientometrics}, 55(2) 317-327.

\end{document}